\documentclass[10pt]{iopart}
\expandafter\let\csname equation*\endcsname\relax
\expandafter\let\csname endequation*\endcsname\relax 
\usepackage{iopams}
\usepackage{epsfig}
\usepackage{amsmath}
\usepackage{amssymb}
\usepackage{graphicx}
\usepackage{cite}
\usepackage{color}

\begin{document}

\def\vecb#1{\boldsymbol{#1}}
\def\ket#1{|#1\rangle}
\def\bra#1{\langle#1|}
\def\scal#1#2{\langle#1|#2\rangle}
\def\matr#1#2#3{\langle#1|#2|#3\rangle}
\def\abs#1{\left\lvert#1\right\rvert}
\def\E#1{\cdot10^{#1}}
\def\ave#1{\langle#1\rangle}
\def\max#1{\{#1\}}
\def\dis#1{\langle\Delta^2#1\rangle}
\def\ske#1{\langle\Delta^3#1\rangle}
\def\={\!=\!}
\def\>{\!>\!}
\def\<{\!<\!}
\def\-{\!-\!}
\def\+{\!+\!}
\def\abs#1{\left|#1\right|}
\def\J{\hat{J}}
\def\b{\hat{b}}
\def\H{\hat{H}}
\def\M{\hat{M}}
\newcommand{\ui}{\mathrm{i}}
\newcommand{\ud}{d}
\def\uvo#1{\lq\lq #1\rq\rq}
\def\ESQPT{\textsc{esqpt}}
\def\ESQPTs{\textsc{esqpt}s}
\def\QPT{\textsc{qpt}}
\def\QPTs{\textsc{qpt}s}


\title[Monodromy in Dicke superradiance]{Monodromy in Dicke superradiance}

\author{Michal Kloc, Pavel Str{\'a}nsk{\'y}, Pavel Cejnar}
\address{Institute of Particle and Nuclear Physics, Faculty of Mathematics and Physics, Charles University, 
  V~Hole{\v s}ovi{\v c}k{\' a}ch 2, 180\,00 Prague, Czech Republic}
\ead{kloc@ipnp.troja.mff.cuni.cz}

\begin{abstract}
We study the focus-focus type of monodromy in an integrable version of the Dicke model.
Classical orbits forming a pinched torus represent analogues of the dynamic superradiance under conditions of a closed system.
Quantum signatures of monodromy appear in lattices of expectation values of various quantities in the Hamiltonian eigenstates and are related to an excited-state quantum phase transition.
We demonstrate the breakdown of these structures with an increasing strength of non-integrable perturbation.
\end{abstract}

\noindent{\it Keywords}: Monodromy, Superradiance, Extended Dicke model


\section{Introduction}

Points of unstable equilibrium of integrable Hamiltonian systems create an obstacle to their fully analytical description \cite{Bat97,Pel11}.
For instance, the single trajectory of a mathematical pendulum that crosses the stationary point of its upper vertical orientation separates two different types of motions in the phase space which are not analytically connectable.
Similar, though more sophisticated singular trajectories are present also in integrable systems with a larger number of degrees of freedom $f$.

A clear example is a spherical pendulum (swings restricted to a spherical surface) with $f\=2$ \cite{Bat97}.
This system is integrable as the component $M$ of angular momentum along the vertical axis is an additional integral of motion besides energy $E$.
After the transformation to action-angle variables, the bundle of $M\=0$ orbits crossing the stationary point on the north pole of the pendulum sphere with energy $E$ equal precisely to the potential energy at that point forms a singular, so-called pinched torus, whose one elementary circle is contracted to a single point.
If approaching the stationary point from two independent directions, the associate momenta linearly contract to zero---we speak about the focus-focus type of singularity \cite{Zou92,Zun97,Pel11}.

The presence of a focus-focus singularity prevents introduction of global action-angle variables valid in the whole $f\=2$ phase space \cite{Bat97,Pel11,Zou92,Zun97,Dui80}.
These variables can be defined on a local level, but in a vicinity of the pinched torus they have some non-trivial topological features.
These become apparent if all tori are imaged in the energy--momentum map $(M,E)$ and if a two-dimensional basis of elementary cycles characterized by angles $(\phi_1,\phi_2)$ is introduced on each torus.
Consider a closed curve encircling the point corresponding to the pinched torus in the energy--momentum map.
A loop along this curve takes us back to the same place, i.e., to the initial torus, but the basis of elementary cycles is altered---linearly transformed by a 2$\times$2 matrix, which is fixed by the number of focus-focus singularities on the pinched torus inside the loop \cite{Bat97,Zun97}.
This situation, when \uvo{once around} does not mean the full return, is captured by the name monodromy \cite{Dui80}.

Monodromy has also specifically quantum signatures \cite{Cus88,Ngo99}.
These can be derived from the application of the semiclassical quantization procedure to integrable systems with singular tori.
It turns out that the joint spectrum (a lattice of energy vs. momentum eigenvalues corresponding to individual Hamiltonian eigenstates) has a defect at the point associated with the pinched torus.
Making a closed loop around this point, one observes a distortion of the lattice elementary cell such that the cell after the loop does not coincide with its initial form. 
The matrix describing the cell transformation is directly related to the classical monodromy matrix deduced from the elementary cycles on tori \cite{Ngo99}.

Effects of quantum monodromy have been identified experimentally in highly excited spectra of some molecules, like H$_2$O and CO$_2$ \cite{Chi07,Cus04}.
More examples and an extensive list of references can be found in Refs.\,\cite{Efs04,Sad10,Zhi11,Dul16}. 
A link has been established between monodromy and so-called excited-state quantum phase transitions \cite{Hei06,Cej06,Lar13}.
These are singularities in the density of energy eigenstates of arbitrary (integrable or non-integrable) systems with any (but preferably low) number of degrees of freedom generated by stationary points of the corresponding classical Hamiltonians \cite{Cap08,Str14}.
For non-degenerate stationary points, the form of the singularity with a given $f$ can be deduced solely from the number of negative Hessian eigenvalues of the Hamiltonian at the stationary point \cite{Str16}. 

In this article we investigate monodromy in an extended Dicke model of single-mode superradiance \cite{Dic54}. 
In particular, we show that the integrable version of the model in its classical limit contains a family of trajectories, which are analogous to the above-mentioned singular orbits of a spherical pendulum.
We describe the defects that appear as a consequence of classical monodromy in quantum lattices of various observables evaluated in the Hamiltonian eigenstates and show a link to a specific excited-state quantum phase transition present in the model. 
In addition, we describe the fate of these singular structures after a gradual breakdown of the system's integrability.
Note that our work represents an extension of Ref.\,\cite{Bab09}, where monodromy in the integrable Dicke model was first studied.

The plan of the paper is as follows: The model is described in Sec.\,\ref{Model} and its integrable version in Sec.\,\ref{Integ}. 
Properties related to classical and quantum monodromy are analyzed in Sec.\,\ref{Mono}.
Breakdown of monodromy under a non-integrable perturbation is studied in Sec.\,\ref{Decay}.
Brief conclusions are given in Sec.\,\ref{Conc}.


\section{Extended Dicke model}
\label{Model}

In 1954, Robert H.\,Dicke predicted an enhancement of spontaneous radiation from atomic or molecular samples caused by a coherent interaction of radiators with the radiation field \cite{Dic54}.
This so-called superradiance can occur if the wavelength of the field is much longer than a typical distance between radiators in the sample. 
The phenomenon has two basic incarnations \cite{Bra05,Kee14}: 
(i) The {\it dynamic superradiance\/} \cite{Dic54}, i.e., a strongly non-exponential, pulse-like decay of the excited sample governed by collective behavior of radiators \cite{Bra05,Kee14,Gro82,Be96}.
This can happen in the form of light emission into free space as well as in a cavity setup with only some discrete field modes present \cite{Fuc16}.
(ii) The {\it equilibrium superradiance\/} \cite{Wan73,Hep73,Ema03}, i.e., the appearance of thermal and quantum phases characterized by a non-zero macroscopic density of radiation in the cavity \cite{Bra05,Kee14}.
Closely related effects have been discussed in nuclear physics \cite{Aue11}, solid-state physics \cite{Con16} and other areas.

Various aspects of superradiance have been tested in laboratory.
The dynamic superradiance as the free-space emission was observed in numerous setups since 1970s (see Ref.\,\cite{Gro82} and references therein).
On the other hand, the observation of the equilibrium superradiance faced a problem of preparing a tunable system with strong atom-field coupling.
A breakthrough was based on the theoretical proposal of Ref.\,\cite{Dim07}, which led to recent experimental realizations of the superradiant phase transition using superfluid Bose gases in an optical cavity \cite{Bau10,Bau11,Kli15} and cavity-assisted Raman transitions  \cite{Bad14}.
These achievements triggered new theoretical efforts aiming at deeper understanding of the superradiance  phenomena.

To illustrate the essence of superradiance, Dicke devised a simple model formulated in terms of a single-mode bosonic field interacting with a chain of two-level atoms, enumerated by $i\=1,\dots,N$ \cite{Dic54}.
While the field quanta are created and annihilated by operators $\b^{\dag}$ and $\b$, the atoms are described by collective quasispin operators $(\J_-\=\J_1\-i\J_2,\J_0\=\J_3,\J_+\=\J_1\+i\J_2)$ composed as sums of Pauli matrices acting in the 2-dimensional Hilbert spaces of individual atoms: $\hat{\vecb{J}}\=\sum_{i=1}^{N}\hat{\vecb{\sigma}}^{(i)}/2$.
We use a slightly extended version of the Dicke Hamiltonian \cite{Bra13,Bas14,Bas16,Klo17}, which can be written in the following form:
\begin{equation}
\H=\omega\ \b^{\dag}\b+\omega_0\J_3+\frac{\lambda}{\sqrt{N}} \left(\b^{\dag}\J_-+\b\J_++\delta\,\b^{\dag}\J_++\delta\,\b\J_-  \right)
\label{H}\,.
\end{equation}
Here, $\omega$ represents a single-boson energy, $\omega_0$ an energy difference between the levels of one atom, and $\lambda$ an overall strength parameter of the atom-field interaction.
We can assume $\lambda\>0$ as the $\lambda\mapsto -\lambda$ conversion is connected with a unitary transformation $(\J_1,\J_2,\J_3)\mapsto(-\J_1,-\J_2,\J_3)$.
The additional parameter $\delta\in[0,1]$ is explained below.

Hamiltonian \eqref{H} can be used as a toy version of the cavity QED.
(Note that the model neglects the term containing the square of the electromagnetic vector potential; for a recent discussion of its role see e.g. Ref.\,\cite{Vie11}.)
In a normal situation, the interaction is written as $\H_{\rm int}\propto\hat{\vecb{E}}\!\cdot\!\hat{\vecb{D}}\propto(\b^\dag\+\b)\J_1$, where $\hat{\vecb{E}}$ is the electric intensity and $\hat{\vecb{D}}$ the atomic dipole-moment matrix element, so $\delta\=1$.
However, for $\lambda\ll\omega,\omega_0$, the terms $\b^{\dag}\J_+$ and $\b\J_-$ give only small contributions to matrix elements and can be neglected \cite{Jay63,Tav68}, so we can set $\delta\=0$.
In this approximation, the model becomes integrable as it conserves the quantity
\begin{equation}
\underbrace{\M}_{M}=\underbrace{\b^{\dag}\b}_{n}+\underbrace{\J_3+j}_{n^*}
\label{M}\,,
\end{equation}
which is the sum of the number of field bosons $n$ and the number of atomic excitation quanta $n^*\=m\+j\leq N^*\equiv 2j$ (symbols under the braces in the above formula stand for eigenvalues of the associated operators and $m$ is an eigenvalue of $\J_3$).
This conservation law follows from a U(1) symmetry of the $\delta\=0$ Hamiltonian under the \lq\lq gauge\rq\rq\ transformation $\b^{\dag}\mapsto e^{i\alpha}\b^{\dag}$, $\hat{\vecb{J}}\mapsto\mathbf{R}(\alpha)\hat{\vecb{J}}$, where $\mathbf{R}(\alpha)$ is the rotation matrix by angle $\alpha$ around axis $z$.
The $\delta\neq 0$ Hamiltonians do not conserve $\M$ but only a parity $\hat{\Pi}=(-)^{\M}$ defining a residual discrete Z(2) symmetry of the system.
Therefore, if the parameter $\delta$ is gradually increased, one goes from the integrable, hence entirely regular regime of dynamics at $\delta\=0$ (so-called Tavis-Cummings limit) to the non-integrable and partly chaotic regime at $\delta\=1$ (Dicke limit).
Although the recent experimental realizations incorporated only the limiting regimes of the model \cite{Bau10,Bau11,Kli15,Bad14}, intermediate values $\delta\in(0,1)$ are in principle also achievable in the general experimental setup of Ref.\,\cite{Dim07}, see \cite{Bha12}.

An essential feature of Hamiltonian \eqref{H} resulting from the required coherence of the atom-field interaction is its strongly collective character, inscribed in the conservation of the squared quasispin $\hat{\vecb{J}}^2$.
This implies a crucial simplification of the analysis since it guarantees that the Hamiltonian acts independently in the subspaces with different $\hat{\vecb{J}}^2$ quantum numbers $j$.
The number $N^*\=2j$, taking values from 0 or 1 (for $N$ even or odd, respectively) to $N\=2j_{\rm max}$, can be considered as the number of {\it active\/} atoms as it measures the maximal excitation energy (in units of $\omega_0$) that can be achieved within the whole atomic ensemble of size $N$; the remaining $N\-N^*\=2(j_{\rm max}\-j)$ atoms form pairs mutually compensating their energies \cite{Cej16}.
A $(2j\+1)$-dimensional subspace with a given value $j$ appears typically in many replicas differing by the permutation symmetry of its states with respect to the exchange of atoms.
The sum of dimensions of all these subspaces exhausts the exponentially increasing dimension $2^N$ of the full Hilbert space of all atomic configurations.
Only the subspace with the highest value $j\=j_{\rm max}$, which contains fully symmetric atomic states, is unique.

The model has two degrees of freedom, $f\=2$: one is connected with the bosonic field, the other with the collective dynamics of atoms.
The classical dynamics was studied by many authors using different techniques, see e.g. Refs.\,\cite{Bab09,Ema03,Bas14,Bha12,Agu92,Bak13}.
The classical limit is achieved if $j\to\infty$, that is $N^*\to\infty$, which implies $j_{\rm max}\to\infty$ and $N\to\infty$.
The result of the limiting process depends on the ratio
\begin{equation}
\gamma=\frac{j}{j_{\rm max}}=\frac{N^*}{N}
\label{rat}
\end{equation}
that can be fixed at a constant value $\gamma\in(0,1]$.
Identifying the model Planck constant $\hbar$ with $(2j)^{-1}\=(N^*)^{-1}$, one can obtain the classical description via the mapping
\begin{eqnarray}
\tfrac{1}{\sqrt{N^*}}\left(\b,\b^{\dag}\right)&\mapsto&
\tfrac{1}{\sqrt{2}}\bigl(x+ip,x-ip\bigr)
\label{map1},\\
\tfrac{1}{N^*}\left(\J_1,\J_2,\J_3\right)&\mapsto&
\left(\sqrt{\tfrac{1}{4}\-z^2}\,\cos\phi,\ \sqrt{\tfrac{1}{4}\-z^2}\,\sin\phi,\ z\right)
\label{map2}\,,
\end{eqnarray}
where $x\in(-\infty,+\infty)$ and $p\in(-\infty,+\infty)$ are associated coordinate and momentum corresponding to the field degree of freedom, while $\phi\in[0,2\pi)$ and $z\in\left[-\frac{1}{2},+\frac{1}{2}\right]$ form the canonically conjugate coordinate-momentum pair for the atomic degree of freedom.
The latter define coordinates ($z$ the latitude projection and $\phi$ the longitude angle) on the Bloch sphere with radius $\frac{1}{2}$.
The scaled Hamiltonian $\H/N^*$ is then mapped to
\begin{equation}
{\cal H}=\omega\,\frac{x^2\+p^2}{2}+\omega_0\,z+
\underbrace{\sqrt{\gamma}\,\lambda}_{\lambda_{\gamma}}\,\sqrt{\frac{1}{2}\-2z^2}
\biggl[(1\+\delta)\ x\cos{\phi}-(1\-\delta)\ p\sin{\phi}\biggr]
\label{Hcl}\,,
\end{equation}
where $\lambda_{\gamma}\in(0,\lambda]$ is a rescaled interaction parameter, which is equal to $\lambda$ for $\gamma\=1$.
The scaled energy values corresponding to Eq.\,\eqref{Hcl} are denoted as ${\cal E}=E/N^*$. 

Eq.\,\eqref{Hcl} enables us to determine quantum critical properties of the atom-field system \cite{Bra13,Bas14,Bas16,Klo17}.
For $\lambda_{\gamma}$ less than a certain critical value $\lambda_{\rm c}$,  the Hamiltonian has a single minimum at $x\=p\=0$ and $z\=-\frac{1}{2}$, implying zero numbers of both atom and field excitation quanta.
At $\lambda_{\gamma}\=\lambda_{\rm c}$, the minimum starts moving to $x,p\neq0$ and $z\>-\frac{1}{2}$. 
This increases separate excitation energies of atoms and field but lets the atom-field interaction reduce the overall ground state energy, which reads as follows:
\begin{equation}
{\cal E}_0=\left\{\begin{array}{ll}
-\frac{\omega_0}{2}&{\rm for\ }\lambda_{\gamma}<\lambda_{\rm c}=\frac{\sqrt{\omega\omega_0}}{1+\delta}\,,\\
-\frac{\omega_0}{2}\frac{\lambda_{\gamma}^4+\lambda_{\rm c}^4}{2\lambda_{\gamma}^2\lambda_{\rm c}^2}&{\rm for\ }\lambda_{\gamma}\geq\lambda_{\rm c}\,.\\
\end{array}\right.
\label{Egr}
\end{equation}
For $\lambda_{\gamma}\>\lambda_{\rm c}$, the system at temperature $T$ lower than a critical temperature $T_{\rm c}=(\omega_0/2){\rm arctanh}^{-1}[\lambda_{\rm c}/\lambda_{\gamma}]^2$ is in the superradiant phase \cite{Bas16,Klo17}.
Moreover, the spectrum of the Hamiltonian eigenvalues with a given ratio $\gamma$ splits into several non-thermal (quantum) phases separated by excited-state quantum phase transitions.
The critical borderlines of quantum phases in the plane $\lambda_{\gamma}\times{\cal E}$ are characterized by distinct singularities in the first derivative of the semiclassical level density \cite{Bra13,Bas14,Bas16,Klo17}.


\section{Classical analysis of the Tavis-Cummings limit}
\label{Integ}

For $\delta\=0$, the classical Hamiltonian in Eq.\,\eqref{Hcl} is integrable.
The scaled integral of motion $\M/N^*$ from Eq.\,\eqref{M} reads as
\begin{equation}
{\cal M}=\frac{x^2\+p^2\+1}{2}+z
\label{Mcl}\,.
\end{equation}
Using the canonical transformation \cite{Klo17}
\begin{eqnarray}
\left(\begin{array}{c}x\\p\end{array}\right)&\mapsto&\left(\begin{array}{c}x'\\p'\end{array}\right)=\left(\begin{array}{cc}\cos\phi&-\sin\phi\\\sin\phi&\cos\phi\end{array}\right)\left(\begin{array}{c}x\\p\end{array}\right)
\label{can1}\,,
\\
\left(\begin{array}{c}z\\\phi\end{array}\right)&\mapsto&\left(\begin{array}{c}z'\\\phi'\end{array}\right)=\left(\begin{array}{c}{\cal M}\-\frac{1}{2}\\\phi\+{\cal M}\-\frac{1}{2}\end{array}\right)
\label{can2}\,,
\end{eqnarray}
the Tavis-Cummings Hamiltonian is converted to the form
\begin{equation}
{\cal H}=\omega_0{\cal M}\-\frac{\omega}{2}+\underbrace{\left(\omega\-\omega_0\right)}_{\Delta\omega}\,\frac{x'^2\+p'^2+1}{2}+\lambda_{\gamma}\,x'\,\sqrt{\frac{1}{2}-2\left({\cal M}\-\frac{x'^2\+p'^2+1}{2}\right)^2}
\label{Hclint}\,,
\end{equation}
which does not depend on angle $\phi'$.
For any fixed value of the quantity ${\cal M}$, the formula \eqref{Hclint} represents dynamics with $f\=1$ effective degree of freedom.

The classical dynamical equations for the transformed Hamiltonian \eqref{Hclint} read as
\begin{eqnarray}
\dot{x}'=\frac{\partial{\cal H}}{\partial p'}&=
+\Delta\omega\,p'+\lambda_{\gamma}\ \frac{2 x'p' z_{\cal M}(r')}{\sqrt{\frac{1}{2}-2z_{\cal M}^2(r')}}
\label{xdot}\,,\\
\dot{p}'=-\frac{\partial{\cal H}}{\partial x'}&=
-\Delta\omega\,x'-\lambda_{\gamma}\,\frac{\frac{1}{2}-2z_{\cal M}^2(r')+2x'^2z_{\cal M}(r')}{\sqrt{\frac{1}{2}-2z_{\cal M}^2(r')}}
\label{pdot}\,,
\end{eqnarray}
where dots stand for the time derivatives and $z_{\cal M}(r')\={\cal M}-(r'^2\+1)/2$ with $r'^2\equiv x'^2\+p'^2$ ($=r^2\equiv x^2\+p^2$).
A return back to the original phase space $(x,p,\phi,z)$ is possible via setting $z\=z_{\cal M}(r')$, see Eq.\,\eqref{Mcl}, and performing an inverse of transformation \,\eqref{can1}, in which the angle $\phi$ is determined via an integration of the dynamical equation  
\begin{equation}
\dot{\phi}=\dot{\phi}'=\frac{\partial{\cal H}}{\partial z'}=\omega_0-\lambda_{\gamma}\,\frac{2x'z_{\cal M}(r')}{\sqrt{\frac{1}{2}-2z_{\cal M}^2(r')}}
\label{phidot}\,.
\end{equation}

\begin{figure}
\begin{flushright}
\includegraphics[width=0.7\linewidth]{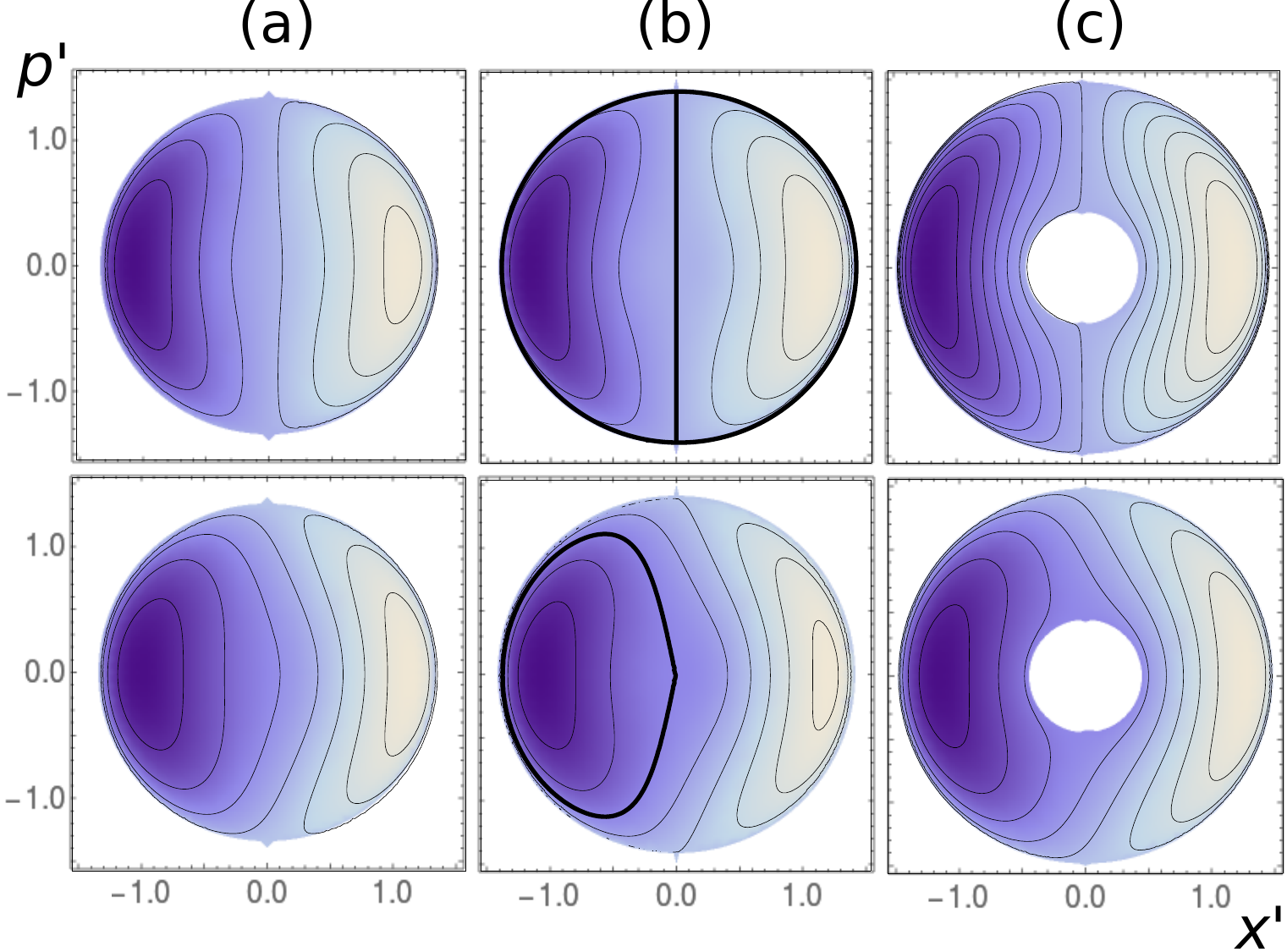}
\end{flushright}
\caption{
(Color online)
Contour plots of the Hamiltonian function \eqref{Hclint} for $\lambda_{\gamma}\!=\!2.5\!>\!\lambda_{\rm c}'$.
Darker areas correspond to the lower values and vice versa.
The upper row depicts the tuned case $\omega\!=\!\omega_0\!=\!1$, the lower row a detuned case $\omega\!=\!2,\omega_0\!=\!1$, the columns correspond to (a) ${\cal M}\!=\!0.9$, (b) ${\cal M}\!=\!1$, and (c) ${\cal M}\!=\!1.1$.
Critical contours passing the point $x'\!=\!p'\!=\!0$ for ${\cal M}\!=\!1$ are marked by thick lines. 
}
\label{F_cont}
\end{figure}

Since the transformation \eqref{can1} conserves radii, $r^2\=r'^2$, the integral of motion \eqref{Mcl} enables one to easily determine for each point $(x',p')$ the corresponding projection $z$ on the atomic Bloch sphere.
This implies a restriction on the available domain in the phase space in the form $r'\in[r'_{\rm min},r'_{\rm max}]$, where
\begin{equation}
r'_{\rm min}=\left\{\begin{array}{ll}
0                         & {\rm for\ }{\cal M}\leq 1\,,\\
\sqrt{2({\cal M}-1)} & {\rm for\ }{\cal M}>1\,,
\end{array}\right.
\qquad
r'_{\rm max}\=\sqrt{2{\cal M}}
\,.
\label{restr}
\end{equation}
The maximum radius $r'_{\rm max}$ defines an outer circle of the available domain which corresponds to the maximum number of field bosons $n\=M$ achieved when all atoms are in the lower state ($n^*\=0$, $z\=-\frac{1}{2}$).
Note that the transformation \eqref{can1} is indeterminate at the outer circle as the angle $\phi$ is irrelevant in the south pole of the Bloch sphere.

The minimum radius $r'_{\rm min}$ in Eq.\,\eqref{restr} results from a minimum number of field bosons $n$ needed to get a given value of $M$ for a maximal atomic excitation $n^*\={\rm Min}(M,N^*)$.
For $M\<N^*$ (${\cal M}\<1$), the minimum number of bosons is zero and the corresponding point $(x',p')\=(0,0)$ is linked to the  Bloch sphere latitude $z={\cal M}\-\frac{1}{2}$, which is less than $+\frac{1}{2}$.
When $M\=N^*$ (${\cal M}\=1$), the $(0,0)$ point gets associated exactly with the north pole $z=+\frac{1}{2}$.
This particular configuration, which represents a state of maximally excited atoms and the field vacuum, will play an essential role in the following. 
Finally, if $M\>N^*$ (${\cal M}\>1$), the number of field bosons $n$ cannot be less than $(M\-N^*)$, so the minimal radius becomes larger than zero, defining an inner circle of the available $(x',p')$ domain.
The whole inner circle corresponds to the north pole of the Bloch sphere, where again the transformation \eqref{can1} becomes undefined.

The contours ${\cal H}\={\cal E}$ of the Hamiltonian \eqref{Hclint} in the available domain of the phase space are shown in Fig.\,\ref{F_cont} for a single interaction strength $\lambda_{\gamma}$ and three values of the integral of motion: ${\cal M}\<1$, ${\cal M}\=1$, and ${\cal M}\>1$, see columns (a), (b), and (c) respectively. 
The upper row depicts the tuned case with $\Delta\omega\=0$, the lower row a detuned case with $\Delta\omega\>0$.
Note that in this paper we assume the detuning hierarchy $\omega\geq\omega_0$, that is $\Delta\omega\geq 0$. 
Properties of the inverse hierarchy $\omega\<\omega_0$ can be derived from the present ones by inverting the whole spectrum upside down, ${\cal H}\mapsto-{\cal H}$ (the ground state becomes the highest state and vice versa), and by applying a reflection transformation $(x',p')\mapsto-(x',p')$ in the phase space.
This converts Hamiltonian \eqref{Hclint} with $\Delta\omega\<0$ into the same form with $\Delta\omega\>0$, up to the constant term which changes its sign.
So all the results discussed below are valid also for the inverse detuning hierarchy, except that the energies need to be suitably transformed. 

The minimum of the transformed Hamiltonian function \eqref{Hclint} determines the lowest energy eigenstate (in the $N^*\!\to\!\infty$ limit) in the selected $M$-subspace of the full Hilbert space (the ground state of the given subspace).
Consider at first the ${\cal M}\<1$ and ${\cal M}\>1$ cases.
These are both characterized by a gradual, smooth evolution of the minimum position and energy with the interaction strength.
Indeed, it can be shown that as $\lambda_{\gamma}$ increases from zero, the minimum moves along the line $p'\=0$,  with $x'$ decreasing below $-r_{\rm min}'$ (for $\Delta\omega\geq 0$) or increasing above $+r_{\rm min}'$ (for $\Delta\omega\<0$).
The scaled energy descends from the initial value ${\cal E}'_0\=\omega_0({\cal M}\-\frac{1}{2})\+\Delta\omega r_{\rm min}^{\prime 2}/2$ taken at $\lambda_{\gamma}\=0$.

A more interesting scenario applies for ${\cal M}\=1$, that is $M\=N^*\=2j$, when the evolution of the spectrum with $\lambda_\gamma$ has a critical character \cite{Klo17,Per11}.
We note that the value ${\cal M}\=1$ demarcates the disc-to-annulus transition of the available phase-space domain and allows any partitioning of $M$ between numbers $n,n^*\in\{0,\dots,N^*\}$.
Let us first assume $\Delta\omega\geq 0$.
In this case, the minimum of the Hamiltonian \eqref{Hclint} stays at $(x',p')\=(0,0)$ up to a certain critical value $\lambda'_{\rm c}$ of the interaction strength, but above this value it starts moving to $x'\<0$ along the $p'\=0$ line.
The system undergoes a ground-state phase transition from a \uvo{non-radiant} state corresponding to maximally excited atoms and no photon to a \uvo{radiant} state with decreasing atomic and increasing field excitations.
The lowest and highest energies ${\cal E}'_0$ and ${\cal E}'_1$ of the classical energy landscape are given by
\begin{eqnarray}
{\cal E}'_0&=\left\{\begin{array}{ll}\frac{\omega_0}{2}                &{\rm for\ }\lambda_{\gamma}\leq\lambda'_{\rm c}=\frac{1}{2}|\Delta\omega|\,,\\
\frac{\omega_0}{2}+\Delta\omega\,s_--2\lambda_{\gamma} s_-\sqrt{1-s_-} &{\rm for\ }\lambda_{\gamma}>\lambda'_{\rm c}\,,
\end{array}\right.
\label{emi}\\
{\cal E}'_1&=\quad\ \tfrac{\omega_0}{2}+\Delta\omega\,s_++2\lambda_{\gamma} s_+\sqrt{1-s_+}
\label{ema}\,,\\
&\qquad\qquad\qquad 
s_{\pm}=\tfrac{2}{3}-\tfrac{2}{9}\tfrac{\lambda_{\rm c}^{\prime 2}}{\lambda_{\gamma}^2}\pm\tfrac{2}{9}\tfrac{\lambda_{\rm c}'}{\lambda_{\gamma}}\sqrt{\tfrac{\lambda_{\rm c}^{\prime 2}}{\lambda_{\gamma}^2}+3}
\nonumber\,.
\end{eqnarray}
The lowest energy ${\cal E}'_0$ shows a discontinuity of its second derivative at $\lambda_\gamma\=\lambda'_{\rm c}$, indicating a second-order ground-state quantum phase transition in the $M\=N^*$ subspace.
Let us stress that the critical strength $\lambda'_{\rm c}$ and minimum energy ${\cal E}'_0$ defined in Eq.\,\eqref{emi} are different from $\lambda_{\rm c}$ and ${\cal E}_0$ related to the global minimum among all $M$-subspaces, see Eq.\,\eqref{Egr}.
In particular, for $\omega_0\!\leq\!\omega\<(3\+\sqrt{8})\omega_0$ we see that $\lambda'_{\rm c}\<\lambda_{\rm c}$, so the present transition can take place deeply in the weak coupling regime, well before the superradiant phase transition of the whole system.
In contrast, the highest energy ${\cal E}'_1$ in Eq.\,\eqref{ema} grows smoothly with $\lambda_\gamma$.
For $\Delta\omega\<0$, the spectrum is inverted, ${\cal E}\to-{\cal E}$, and shifted up by $\omega_0$, so the non-analytic evolution affects on the contrary the highest-energy state, while the ground state evolves smoothly.

Regardless of the detuning hierarchy $\Delta\omega\!\geq\!0$ or $\Delta\omega\<0$, the energy value 
\begin{equation}
{\cal E}'_{\rm c}=\frac{\omega_0}{2}
\qquad 
{\rm for\ }\lambda_{\gamma}>\lambda'_{\rm c}
\label{ece}
\end{equation}
demarcates a point of unstable equilibrium of the system---an inflection-like point of the energy landscape \eqref{Hclint} present above the critical interaction strength $\lambda'_{\rm c}$ from Eq.\,\eqref{emi}.
It implies an excited-state quantum phase transition in the $M\=N^*$ subspace: at the critical energy \eqref{ece} the semiclassical density of levels in this subspace shows a logarithmic divergence \cite{Klo17,Per11}. 
This is an extreme form of spectral singularity resulting from the fact that the classical Hamiltonian \eqref{Hclint} has just a single effective degree of freedom \cite{Str16}.
In contrast, the entire system with $f\=2$, governed by the Hamiltonian \eqref{Hcl}, exhibits at ${\cal E}'_{\rm c}$ a non-degenerate stationary point of ${\cal H}$ with two positive and two negative Hessian eigenvalues, so the full energy spectrum of all-$M$ levels shows a downward jump in the first derivative of the level density \cite{Str16,Klo17}.
Moreover, as we will see below, the point $({\cal M},{\cal E})\!=\!(1,{\cal E}'_{\rm c})$ corresponds to a pinched torus of the focus-focus monodromy.


\section{Classical and quantum monodromy}
\label{Mono}

\begin{figure}
\begin{flushright}
\includegraphics[width=\linewidth]{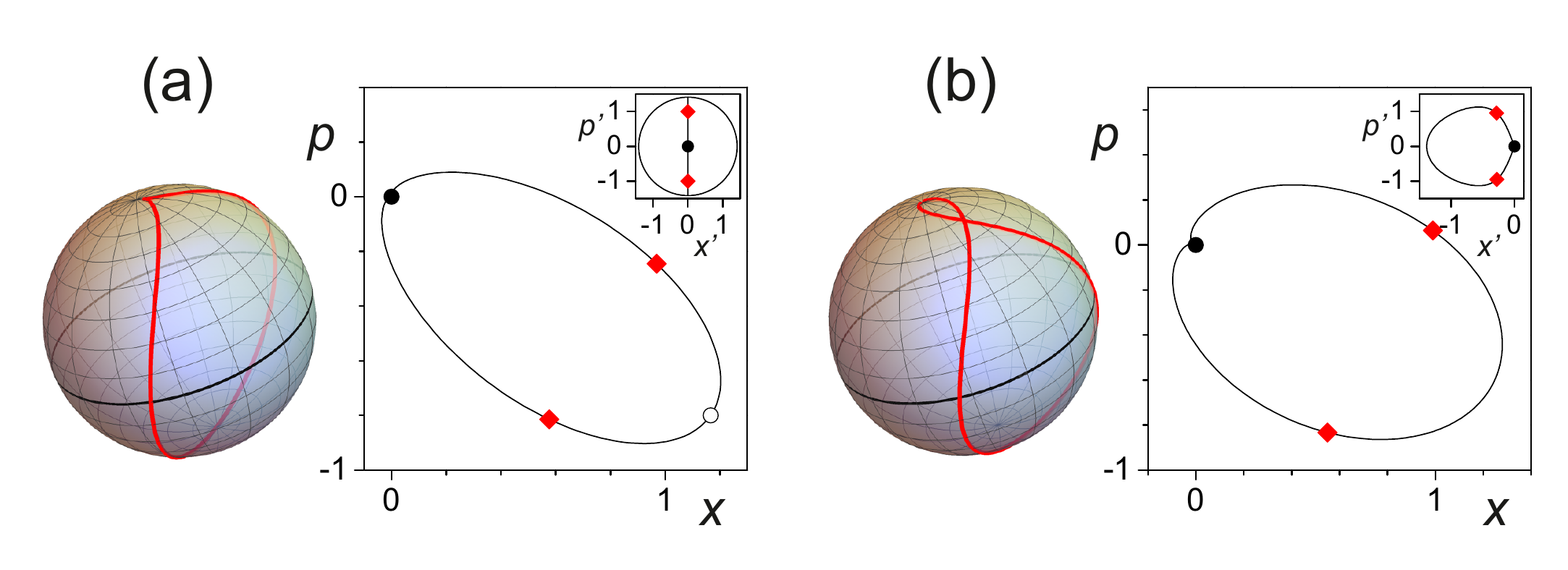}
\end{flushright}
\caption{
(Color online)
Classical orbits crossing the stationary point $(x',p')\!=\!(0,0)$ of Hamiltonian \eqref{Hclint} with $({\cal M},{\cal E})\!=\!(1,{\cal E}'_{\rm c})$ in the (a) tuned and (b) detuned cases. 
The model parameters are as in Fig.\,\ref{F_cont}.
The orbits in $(x',p')$ are depicted in the insets, the main panels show the associated motions on the atomic Bloch sphere $(\phi,z)$ and in the bosonic phase space $(x,p)$. 
Crossings of the north or south pole of the Bloch sphere are coordinated with passages through the full or open bullets in the bosonic space (the south pole is visited only in the tuned case), the equator transits correspond to the diamonds.
Note that the asymptotic spiral motions around the north pole and $(x,p)\!=\!(0,0)$ are under the resolution scale.
}
\label{F_sph}
\end{figure}

Figure~\ref{F_sph} depicts two trajectories with $({\cal M},{\cal E})\=(1,{\cal E}'_{\rm c})$ in the original phase space $(x,p,\phi,z)$, with the atomic variables represented on the Bloch sphere.
Panels (a) and (b) show the tuned $\Delta\omega\=0$ and detuned $\Delta\omega\>0$ cases, respectively.
If drawn in the transformed phase space (see the insets), both orbits coincide with the contours crossing the unstable stationary point $(x',p')\=(0,0)$, see the thick curves in Fig.\,\ref{F_cont}(b). 
The motions along the curves in both atomic and bosonic parts of the phase space in Fig.\,\ref{F_sph} are correlated, so we mark the points on the $(x,p)$ orbit that are on the $(\phi,z)$ orbit synchronized with the pole and equator crossings.
We see that the detuned trajectory crosses only the north pole of the Bloch sphere, while the tuned one goes via both poles. 
Because the representation in terms of transformed variables $(x',p')$ becomes invalid on the outer circle of the available domain, the whole outer segment of the tuned trajectory in the inset of Fig.\,\ref{F_sph}(a) is mapped to a single point in the original space $(x,p)$ shown in the main image.

The north pole of the atomic Bloch sphere represents a special point of both $({\cal M},{\cal E})\=(1,{\cal E}'_{\rm c})$ orbits in Fig.\,\ref{F_sph} as its crossing requires {\em infinite time}.
Time relations for these particular orbits can be most easily deduced in the tuned case, when $x'\=0$ (except the outer circle), so that Eq.\,\eqref{pdot} reduces to
\begin{equation}
\dot{p}'=-\sqrt{2}\,\lambda_{\gamma}\,\sqrt{\frac{1}{4}-z^2_{{\cal M}}(r')}=-\frac{\lambda_{\gamma}}{\sqrt{2}}\,\sqrt{p'^2(2-p'^2)}
\label{pdot2}
\end{equation}
and Eq.\,\eqref{phidot} yields $\dot{\phi}\=\omega_0$.
We see that the momentum derivative vanishes for $p'\=0$ (the north pole, $z\=+\frac{1}{2}$), but also for $p'\=\pm\sqrt{2}$ (the south pole, $z\=-\frac{1}{2}$).
However, the latter \uvo{stationary} points are false ones as they lie on the outer circle where the transformation \eqref{can1} is indeterminate. 
It can be shown that the crossing of these points takes a finite time, in contrast to the real stationary point at $(x',p')\=(0,0)$.
If $\vecb{\rho}\equiv(\rho_1,\rho_2)$ denotes a projection of the $(\phi,z)$ point of the Bloch sphere onto the equator plane, the evolution close to the north pole is approximated by
\begin{equation}
\vecb{\rho}\ \propto\  e^{\pm\lambda_{\gamma}t}(\cos\omega_0t,\sin\omega_0 t)
\label{spiral}\,,
\end{equation}
where time $t$ is counted so that $\vecb{\rho}\=(\rho_1,0)$ at $t\=0$.
Eq.\,\eqref{spiral} defines a spiral winding in the inward or outward direction around the focus $\vecb{\rho}\=0$.
Note that a similar whirl appears also in the bosonic phase space around $(x,p)\=(0,0)$, but these structures are so tiny that they cannot be seen in  Fig.\,\ref{F_sph}.
The detuned trajectory does not cross the south pole, so it avoids problems with the outer circle, but the time relations at the north pole are similar as in the tuned case.

The single critical trajectory shown in either panel (a) or (b) of Fig.\,\ref{F_sph} is not isolated.
It belongs to the respective infinite bundle of orbits differing by angle $\phi_0\in[0,2\pi)$ at the initial point (e.g. at the equator of the Bloch sphere).
The elementary cycles connected with varying $\phi$ shrink to a single point as the orbit approaches the north pole.
We infer that the critical orbits in either $\Delta\omega\>0$ or $\Delta\omega\=0$ case form a pinched torus, so the system exhibits a focus-focus type of monodromy.

There is some similarity between the $({\cal M},{\cal E})\=(1,{\cal E}'_{\rm c},)$ orbits in the present closed cavity system and  the dynamic superradiance phenomenon in open systems \cite{Kee14,Fuc16}.
Indeed, the photon emission/absorption rate is given by
\begin{equation}
\dot{n}=N^*\left(x'\dot{x}'+p'\dot{p}'\right)
\label{ndot}
\,,
\end{equation}
which in the tuned case can be directly evaluated from Eq.\,\eqref{pdot2}.
The largest time derivative takes place at $p'\=1$, which corresponds to the equator of the atomic Bloch sphere ($z\=0$), and a similar conclusion, based on Eqs.\,\eqref{xdot} and  \eqref{pdot}, is valid also in the detuned case.
So the critical orbits describe a non-exponential decay of a fully excited atomic ensemble followed by complete re-absorption of emitted photons, a process which is infinitesimally slow when $(n,n^*)\approx(0,N^*)$, fast when $(n,n^*)\approx(N^*/2,N^*/2)$, and slow again when $(n,n^*)\approx(N^*,0)$.
This resembles the pulse-like decay process associated with dynamic superradiance, although Eq.\,\eqref{ndot} implies a strictly linear scaling of the irradiation peak with $N^*$, which is in contrast to the non-linear scaling valid for the free-space superradiance \cite{Gro82}.

\begin{figure}
\begin{flushright}
\includegraphics[width=0.8\linewidth]{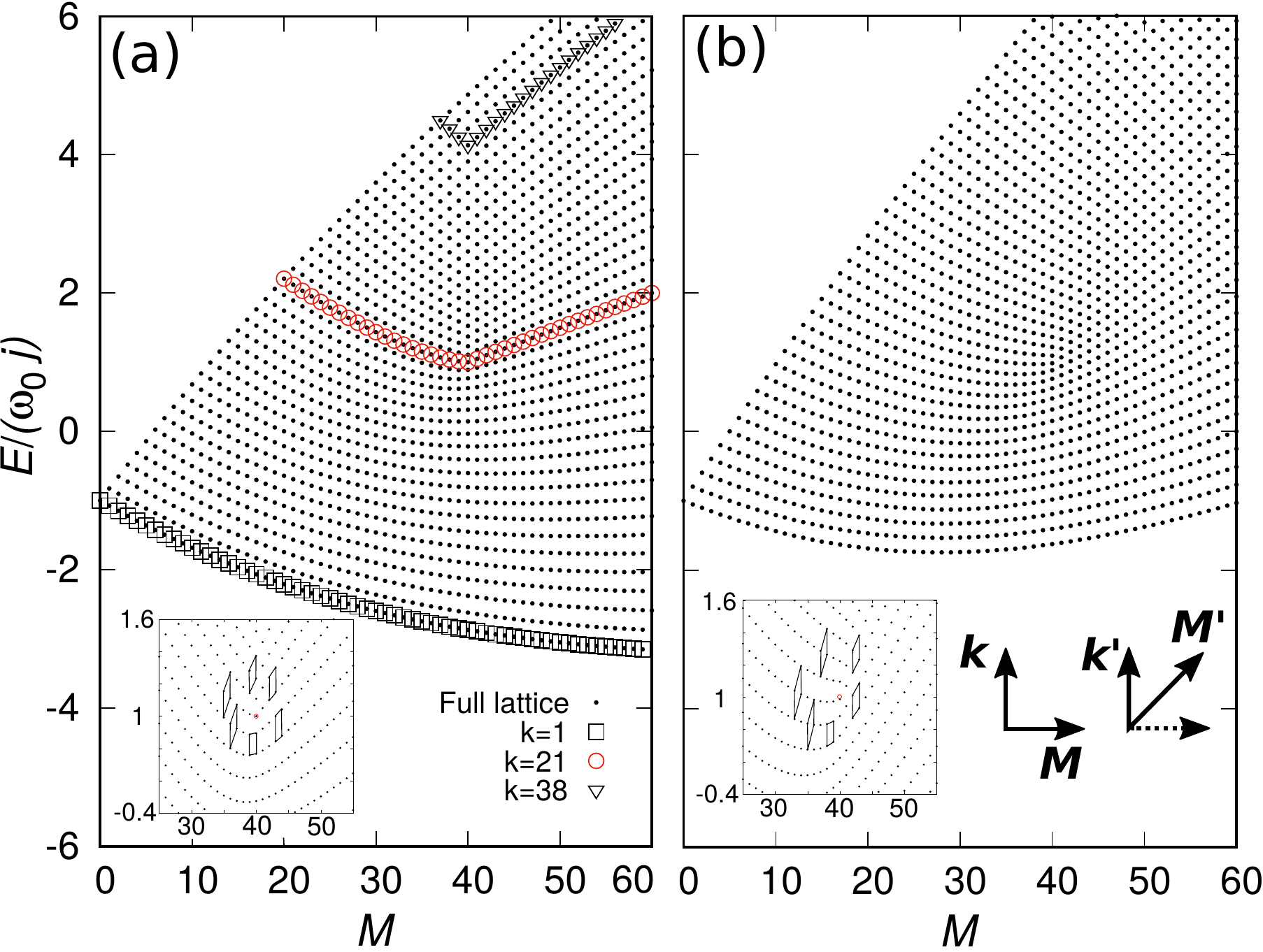}
\end{flushright}
\caption{
(Color online) Quantum energy-momentum maps of the Hamiltonian \eqref{H} with $\lambda\!=\!2.5$, $\delta\!=\!0$ and $N\!=\!2j\!=\!40$. 
The $\omega\!=\!\omega_0\!=\!1$ and $\omega\!=\!2$, $\omega_0\!=\!1$ lattices of eigenstates are in panels (a) and (b), respectively.
The highlighted chains of points correspond to the eigenstates with the same principal quantum number $k$ and variable $M$.
The insets show a transformation of the elementary lattice cell after a closed loop around the monodromy point.
}
\label{F_EM}
\end{figure}

Let us investigate quantum signatures of monodromy.
Fig.\,\ref{F_EM} shows quantum energy-momentum maps \cite{Efs04,Sad10,Zhi11,Dul16}, which are lattices of individual quantum energies $E_k$ (where $k$ stands for a principal quantum number simply enumerating energy eigenvalues) versus the quantum number $M$.
The lattices corresponding to tuned $\Delta\omega\=0$ and detuned $\Delta\omega\>0$ spectra are shown in panels (a) and (b), respectively.
At each point $(M,E_k)$ of the lattice we can construct an elementary cell, that is a rectangle defined by \uvo{horizontal} and vertical basis vectors $\vecb{M}\equiv(1,E_k(M\+1)\-E_k(M))$ and $\vecb{k}\equiv(0,E_{k+1}(M)\-E_k(M))$.
The lattice has a defect at the monodromy point $(M,E)\=N^*(1,{\cal E}'_{\rm c})\=(2j,\omega_0 j)$.
Following a closed loop around this point, as shown in the insets of Fig.\,\ref{F_EM}, the basis vectors undergo a transformation 
\begin{equation}
\left(\begin{array}{c}\vecb{M}'\\\vecb{k'}\end{array}\right)=\underbrace{\left(\begin{array}{cc}1&1\\0&1\end{array}\right)}_{\vecb{\mu}^{\rm T}}\left(\begin{array}{c}\vecb{M}\\\vecb{k}\end{array}\right)
\label{mon}\,,
\end{equation}
where $\vecb{\mu}^{\rm T}$ is a transpose of the classical monodromy matrix $\vecb{\mu}$ describing the transformation of elementary cycles on tori after a loop around the pinched torus containing a single focus-focus singularity\cite{Ngo99}.

The lattice defect in the energy-momentum map can also be manifested by connecting the sequences of points $E_k(M)$ with fixed $k$ and variable $M$.
Three such sequences are highlighted in Fig.\,\ref{F_EM}(a).
While all sequences below the monodromy point show a smooth bend, suggesting a quadratic dependence of energy on $M$, those above the monodromy point exhibit a sharp break, consistent with a linear type of the dependence.
This is in general related to different nature of excitations (e.g. rotational and vibrational in molecular realizations of monodromy) below and above the critical energy \cite{Chi07,Cus04,Efs04,Sad10}.
We stress that in our system, the monodromy energy \eqref{ece} coincides with the critical borderline for an excited-state quantum phase transition present in the $M\=N^*$ subspace \cite{Per11,Klo17}.
This coincidence is analogous to the cases reported in Refs.\,\cite{Hei06,Cej06,Lar13} and can be anticipated as rather common also in other $f\=2$ integrable realizations of excited-state quantum phase transitions.

\begin{figure}
\begin{flushright}
\includegraphics[width=\linewidth]{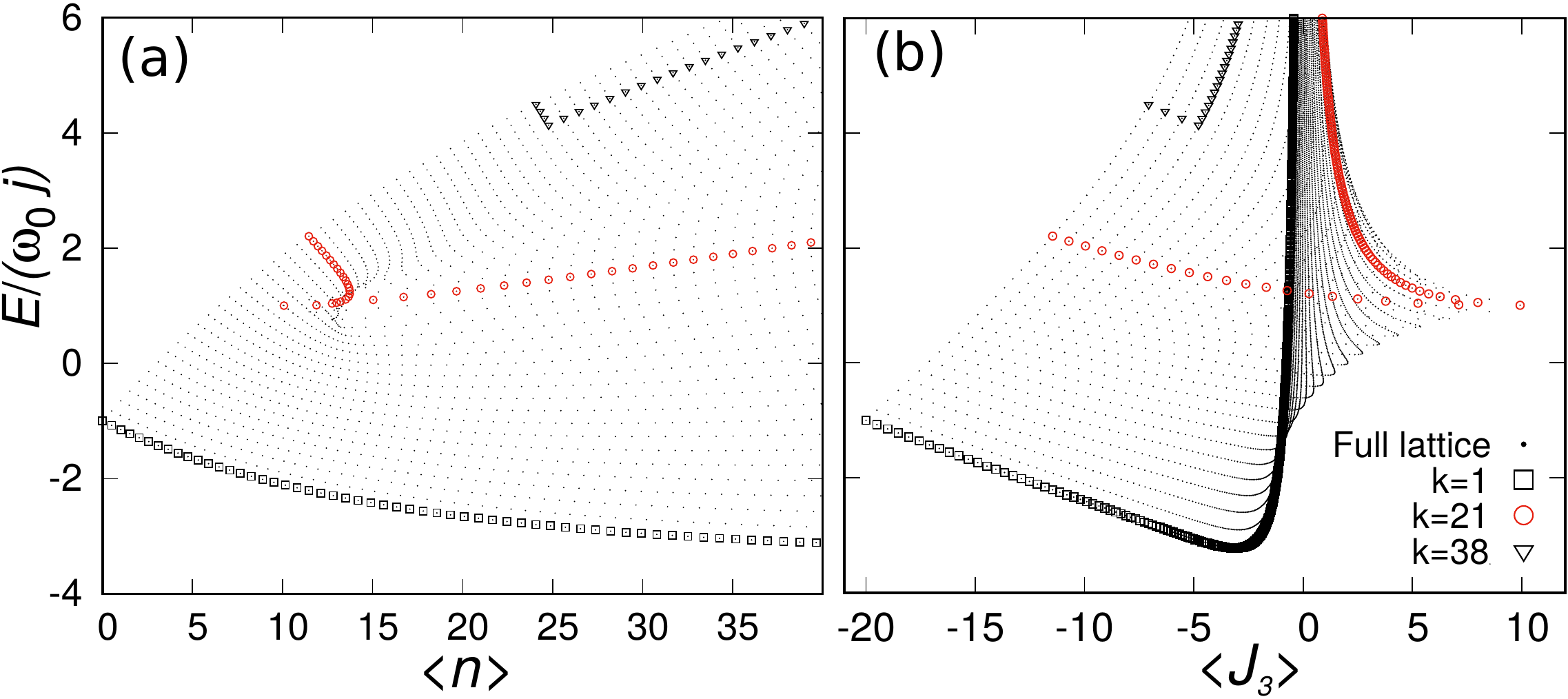}
\end{flushright}
\caption{
(Color online) Peres lattices of observables $\hat{n}\!=\!\b^{\dagger}\b$ and $\J_3$ for Hamiltonian \eqref{H} with the same parameter values as in Fig.\,\ref{F_EM}(a).
The chains of points with the same principal quantum number are again highlighted.
}
\label{F_Per}
\end{figure}

Quantum energy-momentum maps display joined spectra of two compatible integrals of motions. 
It means that both quantities are sharply determined in each eigenstate. 
However, one can also create eigenstate lattices with the abscissa capturing just an {\em expectation value\/} of an arbitrary---i.e., generally not conserved---quantum observable. 
Since this representation of a general spectrum, which is not restricted to integrable systems, was proposed by A. Peres \cite{Per84}, we call it a Peres lattice.
It is useful in the visualization of quantum chaos in mixed systems, in which the lattice contains both ordered and disordered domains, see e.g. Refs.\,\cite{Bas14,Klo17,Str09}.
Two Peres lattices of the integrable Dicke model with $\Delta\omega\=0$ are depicted in Fig.\,\ref{F_Per}.
The first one, in panel (a), shows expectation values $\ave{n}_k$ of the number of field bosons in the eigenstates with energy $E_k$, the second lattice in panel (b) shows in the same manner the expectation values $\ave{J_3}_k$ of the quasispin $z$-projection.
Note that  if the values $\ave{n}_k$ and $\ave{J_3}_k$ at each point are summed, one would obtain precisely the lattice in Fig.\,\ref{F_EM}(a) shifted by a constant $j$, see Eq.\,\eqref{M}.

We observe that both lattices in Fig.\,\ref{F_Per} exhibit apparent singularities at the monodromy point.
However, the topologies of these singularities differ considerably from each other and from that in the energy-momentum map.
Taking this observation the other way round, we can conclude that appearance of various defects in arbitrary Peres lattices may serve as a useful heuristic indicator of monodromy in a general (otherwise unknown) quantum system.

An important task related to a possible experimental verification of quantum monodromy is to identify its signatures in the structure of eigenstates and in the time evolution.
The key observation in the present system is that the $\lambda_\gamma\>\lambda'_{\rm c}$ eigenstates with $M\=N^*$ in a vicinity of the monodromy energy $E\=N^*{\cal E}'_{\rm c}$ exhibit a very large overlap with the unperturbed ground state $(n,n^*)\=(0,N^*)$.
This localization becomes singular for $N^*\!\to\!\infty$ and follows from the diverging time spent by the classical trajectory in an infinitesimal vicinity of the $(x',p')\=(0,0)$ stationary point.
An analogous behavior is observed also in other $f\=2$ systems with the focus-focus type of monodromy \cite{Cej06,San15} as well as in $f\=1$ systems with a local maximum of the Hamiltonian \cite{Car92,Ley05}.
In the present case, the localization results in a sharp local decrease of the atom-field entanglement entropy \cite{Klo17} and has specific consequences for the dynamics of relaxation processes following a quantum quench \cite{Per11,San15}.
In particular, a quench from the $\lambda\<\lambda'_{\rm c}$ ground state to the $\lambda_\gamma\>\lambda'_{\rm c}$ critical region results in a slow decay of the initial state (due to its large overlap with the eigenstates in the critical region) \cite{San15}, while a quench from the $\lambda\>\lambda'_{\rm c}$ side leads, on contrary, to an immediate decay of the initial state and its weak re-occurrences \cite{Per11}.
A more detailed analysis of the quantum quench dynamics in the extended Dicke model is presently a subject of our study.


\section{Decay of monodromy}
\label{Decay}

Monodromy in its original form is restricted solely to integrable systems.
However, an extension of this concept was proposed to softly chaotic systems, in which the singular torus survives the perturbation in the sense of the Kolmorogov-Arnold-Moser theorem \cite{Bro07}.
The present model enables us to study the \uvo{fate} of monodromy in the non-integrable regime explicitly, by setting a non-zero value of parameter $\delta$ in Hamiltonian \eqref{H}.

\begin{figure}
\begin{flushright}
\includegraphics[width=\linewidth]{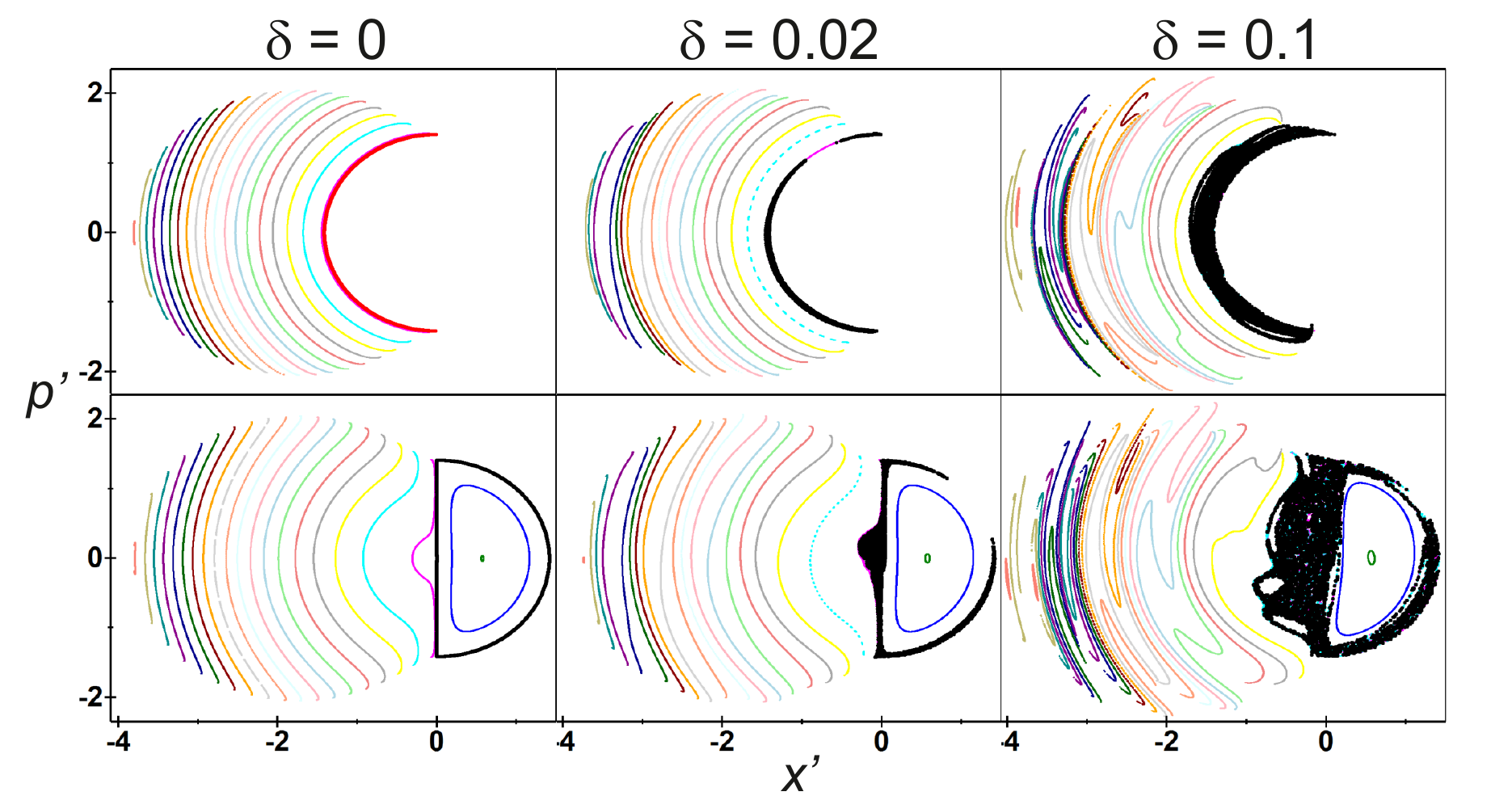}
\end{flushright}
\caption{
(Color online) Poincar{\' e} sections showing passages of 21 classical orbits through the plane $\phi'\!=\!0$ at energy ${\cal E}\!=\!{\cal E}'_{\rm c}$ for Hamiltonian \eqref{Hcl} with $\omega\!=\!\omega_0\!=\!1$ and $\lambda_{\gamma}\!=\!2.5$.
The three columns correspond to the indicated values of parameter $\delta$. 
The upper row collects passages of orbits from negative to positive $\phi'$ values, the lower row passages in the opposite direction.
Colors (online) distinguish individual orbits, black denotes the orbit with an average value $\ave{{\cal M}}$ very close to unity (the value ${\cal M}\!=\!1$ corresponds to the pinched torus for $\delta\!=\!0$).
}
\label{F_Poi1}
\end{figure}

\begin{figure}
\begin{flushright}
\includegraphics[width=\linewidth]{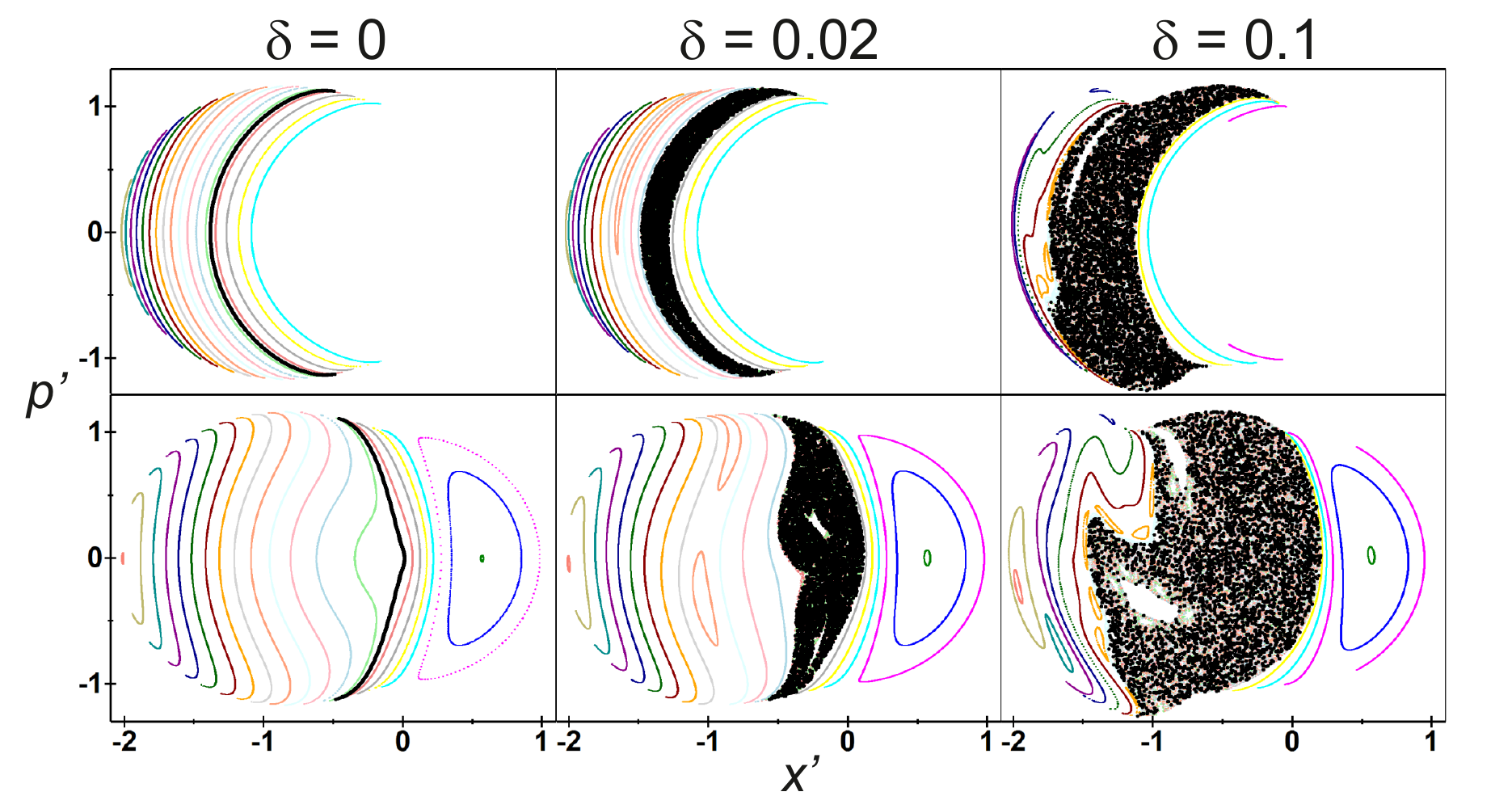}
\end{flushright}
\caption{
(Color online) The same as in Fig.\,\ref{F_Poi1}, but for $\omega\!=\!2,\omega_0\!=\!1$.
}
\label{F_Poi2}
\end{figure}

Poincar{\'e} sections for a sample of classical trajectories having precisely the energy ${\cal E}\={\cal E}'_{\rm c}$ of the monodromy point, and their evolution with increasing $\delta$, are seen in Figs.\,\ref{F_Poi1} and \ref{F_Poi2}.
The dynamics was calculated from the general Hamiltonian \eqref{Hcl} with the coordinates and momenta transformed according to Eqs.\,\eqref{can1} and \eqref{can2}.  
The figures show multiple passages of individual orbits through the plane $\phi'\=0$ in the phase space for the transformed Hamiltonian \eqref{Hclint} with 
$\Delta\omega\=0$ (Fig.\,\ref{F_Poi1}) or $\Delta\omega\>0$ (Fig.\,\ref{F_Poi2}).
The perturbation strength $\delta$ grows from the left column to the right.
The upper and lower rows separate two different directions of the orbit passage through the plane of the section.
If forward and backward segments of the same orbit were plotted in the same figure, they would form a closed curve, but the curves corresponding to different orbits would cross each other.
This behavior (which is in contrast to common Hamiltonians with a quadratic dependence on momenta) is due to a non-trivial (often non-monotonous) evolution of angle $\phi$, see Eq.\,\eqref{phidot}.

The leftmost panels of Figs.\,\ref{F_Poi1} and \ref{F_Poi2}, which correspond to the integrable regime, carry complementary information to previously discussed Fig.\,\ref{F_cont}.
The previous figure displayed different ${\cal E}$ contours of the Hamiltonian \eqref{Hclint} with fixed ${\cal M}$, while the present figures represent $(x',p')$ solutions of Eq.\,\eqref{Hclint} with fixed ${\cal E}$ and different ${\cal M}$.
In all these figures, we can identify the critical orbits with $({\cal M},{\cal E})\=(1,{\cal E}'_{\rm c})$ that belong to the pinched torus.
As $\delta$ increases from zero (in the middle and right panels of Figs.\,\ref{F_Poi1} and \ref{F_Poi2}), the quantity ${\cal M}$ is no more conserved and the orbits can be characterized only by time averages $\ave{{\cal M}}$.
Therefore, if focusing on the ${\cal E}\={\cal E}'_{\rm c}$ orbits with $\ave{{\cal M}}\approx 1$ (passages of these orbits are plotted by the darkest, black shade), we pursue the evolution of the pinched torus and its close neighbors in the perturbed system.

It is clear from Figs.\,\ref{F_Poi1} and \ref{F_Poi2} that the $\ave{{\cal M}}\!\approx\!1$ orbits become chaotic at the earliest stage of the system's perturbation.
Already in the middle panels of both figures, that is at a very moderate value of $\delta$, these orbits generate a distinct quasi-ergodic domain in the phase space (bounded cloud of random crossings).
This domain further grows with increasing $\delta$.
Note that the observed instability of the pinched torus results from the unstable character of the $(x',p')\=(0,0)$ stationary point passed by its orbits.

\begin{figure}
\begin{flushright}
\includegraphics[width=\linewidth]{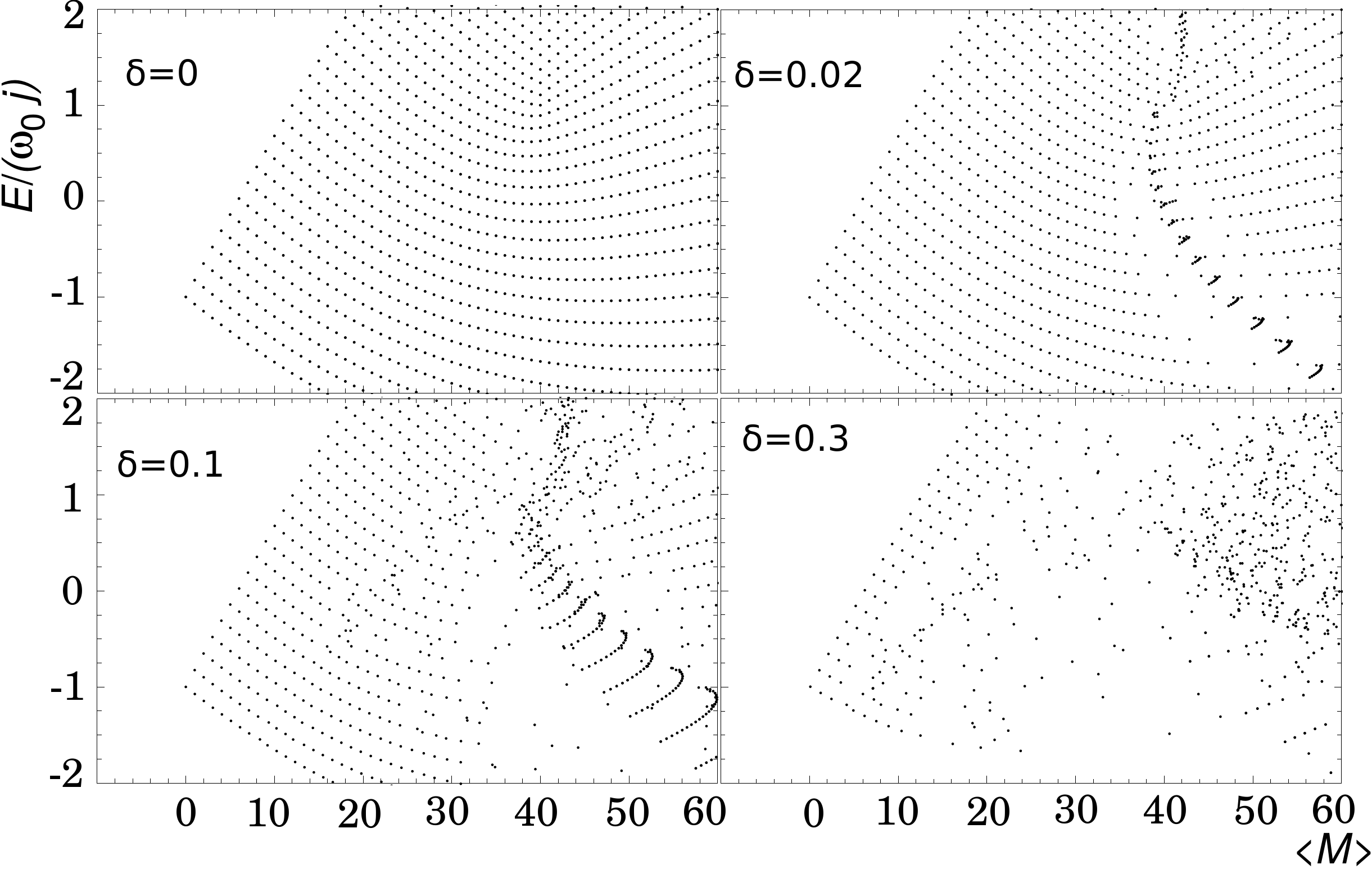}
\end{flushright}
\caption{
Breakdown of the quantum energy-momentum lattice from Fig.\,\ref{F_EM}(a)
with increasing perturbation $\delta$.

}
\label{F_Dec}
\end{figure}

What is the quantum counterpart of the above classical scenario?
Metamorphoses of the quantum energy-momentum map of a tuned system with increasing perturbation strength are depicted in Fig.\,\ref{F_Dec}.
In the upper-left panel we see the lattice corresponding to the $\delta\=0$ Hamiltonian; it is identical with the lattice in panel (a) of Fig.\,\ref{F_EM}. 
The other three panels show what happens if $\delta$ is increased.
In these cases, $M$ on the horizontal axis represents only a quantum expectation value $\ave{M}$ in individual eigenstates.
The first two non-zero values of $\delta$ in Fig.\,\ref{F_Dec} were chosen the same as in the Poincar{\'e} section figures---at these perturbation strengths we observe initial stages of the lattice destruction.
The fourth value of $\delta$ in Fig.\,\ref{F_Dec} is larger, and we already find a considerable part of the lattice being completely messed up.
This agrees with the original use of Peres lattices for visualization of chaos in quantum systems \cite{Per84}.

As seen in Fig.\,\ref{F_Dec}, the point defect defining the quantum monodromy in the energy-momentum lattice is destroyed already with the weakest non-integrable perturbation of the Hamiltonian.
It happens to be right at the center of a large break that splits the spectrum in the vertical direction.
This is not an accident.
The break starts developing along a line where the energy spacing $\Delta E\=E_{k+1}\-E_k$ between neighboring levels is minimal.
As follows from basic perturbation theory, for these states the perturbation efficiency is particularly large due to small energy denominators in the corresponding expressions.
The monodromy point naturally belongs to this line.
Indeed, the simple semiclassical relation $\Delta E=2\pi\hbar/\tau$, connecting the energy spacing $\Delta E$  in an $f\=1$ system with the period $\tau$ of classical motions at the corresponding energy, indicates that the $\tau\to\infty$ orbits on the pinched torus generate very dense, $\Delta E\to 0$, regions of quantum spectra.
In these parts, any generic perturbation of the Hamiltonian results in a fast level repulsion.
The same mechanism of chaos proliferation was observed also in other quantum systems \cite{Str09}. 
So we may conclude that in a typical situation spectral defects of the present type do not survive too long in the non-integrable regime.


\section{Conclusions}
\label{Conc}

We studied monodromy of the focus-focus type  in the integrable Tavis-Cummings limit of the Dicke model.
We showed that the pinched torus in the phase space is formed by orbits with total energy $E$ equal to the energy of a fully excited atomic ensemble and momentum $M$ (the total number of atomic and field excitation quanta) equal to the number of excitable atoms.
These orbits, which are partly similar to critical orbits of a spherical pendulum, represent an analog of the dynamic superradiance phenomenon under circumstances of a strictly closed system.
In particular, the initial state of maximally excited atoms and no field in the cavity becomes a very slowly decaying configuration (a point of unstable equilibrium in the infinite size limit), while the fast decay takes place on a halfway to the full atomic de-excitation.
This resembles the superradiant peak known from the open Dicke systems, although the scaling with the number of atoms is only linear in the closed case.

On the quantum level, monodromy shows up as a point defect in the discrete energy-momentum map, and as a singularity in other Peres lattices.
Quantum signatures of monodromy are closely related to an excited-state quantum phase transition in the critical $M$-subspace of Hamiltonian eigenstates, in particular to a sharp local increase (logarithmic divergence in the infinite-size limit) of the density of states within this subset.
On the other hand, the total density of states in all $M$-subspaces exhibits only a discontinuity of its first derivative at the corresponding energy.
We anticipate that this behavior is common to all $f\=2$ systems with the focus-focus singularity as their Hamiltonians close to the singularity can be cast in a locally quadratic, thus separable form with two positive and two negative Hessian eigenvalues (e.g., as a Hamiltonian with a quadratic kinetic term near a quadratic potential maximum). 
Dynamical consequences of these phenomena are subject of ongoing research.

We have shown that classical and quantum signatures of monodromy in our model disappear already with a very weak perturbation of the system.
We anticipate that fragility is a rather common property of the present type of monodromy as the underlying classical stationary points are unstable (therefore inclined to chaotic dynamics) and imply infinite-period orbits (which are connected with dense, hence vulnerable parts of quantum spectra).

\section*{Acknowledgments}
In memory of Tobias Brandes, whose gentle guidance and encouragement are painfully missed.
This work was supported by the Czech Science Foundation under project no. P203-13-07117S.


\end{document}